# Near Field optical image of a gold surface: a luminescence study

A. Merlen[1*], J. Plathier[2], A. Ruediger[2*]

1 IM2NP, UMR 7334, Site de l'Université de Toulon, 83957 La Garde Cedex (France)

2 INRS, Centre Énergie Matériaux Télécommunications, 1650, boulevard Lionel-Boulet

Varennes (Québec) J3X 1S2 (Canada)

* Corresponding authors: merlen@univ-tln.fr, ruediger@emt.inrs.ca

**Abstract:** This paper addresses recent experimental findings about luminescence of a gold tip in near-field interaction with a gold surface. Our electrochemically etched gold tips show a typical, intrinsic luminescence that we exploit to track the plasmon resonance modeled by a Lorentzian oscillator. Our experimental device is based on a spectrometer optically coupled to an atomic force microscope used in tuning fork mode. Our measurements provide evidence of a strong optical coupling between the tip and the surface. We demonstrate that this coupling strongly affects the luminescence (Intensity, wavelength and FHWM) as a function of the tip position in 2D maps. The fluctuation of these parameters is directly related to the plasmonic properties of the gold surface and is used to qualify the optical near field enhancement (which subsequently plays the predominant role in surface enhanced spectroscopies) with a very high spatial resolution (typically around 20 nm). We compare these findings to the independently recorded near-field scattered elastic Rayleigh signal.

**Introduction**

Noble metal surfaces play a predominant role in plasmon-based surface enhanced spectroscopy techniques. Their investigation requires characterization tools with a very high spatial resolution, traditionally reserved to electron microscopy techniques. In the particular case of metallic nanostructures the strong near field electromagnetic enhancement has a typical range of a few nanometers[1]. As it plays a key role in surface enhanced spectroscopy[2,3] (SERS, SEF SEIRA etc.), the determination of its spatial distribution around metallic nanostructures is required to achieve a better understanding of the mechanisms involved in any of these techniques. For this reason a variety of experimental methods[4] have been adapted for the investigation of the near field. Some use electronic microscopy (e.g. EELS[5], Cathodolumuniscence[6]). This technique has demonstrated several benefits (high resolution for both spectral and spatial measurements) but it has also some major limitations such as the requirement for conductive samples, high vacuum environments, transparency for electron beam for EELS measurements.In addition the exact nature of the interaction between electrons and plasmons is still under debate even if its understanding has made substantial progresses recently[7]. For these reasons, optical measurements are generally privileged. However, it was not until the end of the 1990s that optical scanning probe techniques bypassed the diffraction limit of classical optical measurements. Scanning near field Optical Microscopy (SNOM) is divided in two main approaches: aperture and apertureless. Aperture SNOM has already demonstrated its efficiency for the study of metallic plasmonic structures[8] but this technique has a weak optical collection leading to a tradeoff between signal strength and spatial resolution. For this reason, apertureless SNOM (sometimes referred to as aSNOM) has been proposed: the probe is a tip (AFM or STM tip) whose role is to transform an undetectable near field signal into a measurable far field one. For the study of plasmonic nanostructures specific tips have been developed[9,10]. The nature of the optical interaction between the tip and the sample is the key point in aSNOM[11,12]. To ensure a strong far field signal, the coupling between the tip and the sample needs to be strong. This is the case with plasmonic tips: the tip-sample interaction can generate highly enhanced and confined (a few nm) near fields leading to a measurable far field signal above the confocal background. High spatial resolution can thus be obtained and measurement down to even single orbital level has already been reported[13,14]. aSNOM has been used to perform measurements of elastic-scattering phenomena and also inelastic scattering such as Raman (Tip Enhanced Raman Spectroscopy[15]), fluorescence[16], second harmonic detection[17], *etc.* The high spatial resolution of aSNOM is thus steadily transforming into a table-top alternative to electron microscopy in the field of plasmionics with the additional virtue of flexible environments. Technically, even liquid media are accessible.

Even aSNOM techniques have several remaining challenges. While the alignment of the tip and the laser focus has been improved through persistent progress in instrumentation, the signal to noise ratio relies on the availability of excellent near field antennae. This is particularly true when measurements are performed on non-transparent samples, which is the case for nanoelectronics and metallic surface. For these samples, optical side illumination and collection is ideal[18] and requires a dedicated AFM head to accommodate a long working distance objective with high optical aperture while at the same time minimizing mechanical loops. For chemical and structural mapping, *i.e.* the acquisition of hyperspectral images, the total time is a relevant factor and scales linearly with the quality factor of the near field antenna. For instance, for a 128 × 128 pixel image, an acquisition of 1 s (which is already rather low for many optical signals, in particular inelastic scattering) it will take more than 4 hours to get the full mapping. At this time-scale, thermal drift of piezoelectric systems needs to be taken into consideration. This drift still remains a limitation in the final possible spatial resolution in SNOM measurements: even if the theoretical resolution can be very high, the final experimental resolution will be fixed by the number of points to limit the total acquisition time.

In most aSNOM studies, the signal originates from the sample itself and is enhanced by the tip. In this configuration, the tip acts like a nanoantenna, which enhances the optical response of the sample. We now introduce the inverse configuration: the optical signal from the tip itself is enhanced by the plasmonic properties of the sample to provide a position-dependent signal: the intensity of the signal is used to map the plasmonic properties of the sample. In such a configuration, the spatial resolution can be better than 10 nm and relies on the radius of the tip. The main challenge of this approach is the realization of a tip with a measureable optical signal. Functionalized tip with fluorescent or Raman active[19] molecules have been proposed[20] but their realization remains problematic and their final size can strongly affects the spatial resolution.

In this article, we therefore introduce an alternative approach: the luminescence from the gold tip itself is strongly enhanced by the interaction with a gold surface. We have developed a reproducible experimental procedure for the preparation of luminescent gold tips with a very small radius (typically a few nm). In such a configuration, a strong signal is measured with short integration times and without compromising the spatial resolution. The final contrast in the optical mappings originates from the variation of the tip surface coupling due to the plasmonic properties of the substrate. Such an approach offers different advantages: a relatively simple optical device without complex post-treatment analysis, high spatial resolution, short acquisition times and limited topographic artifacts compared to the elastically scattered signal.

**Method**

The optical near filed measurements are performed with an AIST-NT OmegaScope 1000 using gold tips and He-Ne laser excitation (632.8 nm, TEM00). A schematic view of this experimental device can be seen in *Figure 1*. The beam power was set to 300 µW and its polarization is aligned parallel to the tip axis. The tip is illuminated using a 0.7 NA objective (Mitutoyo M Plan Apo 100x) mounted on a piezoelectric stage oriented at 65° form the tip axis. The piezoelectric objective scanner has a range of 15 µm along the optical axis and 30 µm on the other directions. That allows for a precise alignment of the tip apex with the focal spot of the objective by directly tracking the luminescence emission of the tip. Optical and topographic maps were performed using the tip in a shear force microscopy mode. The optical signal was collected by the same objective. Elastically back-scattered light is simultaneously detected with a Photomultiplier whereas luminescence is analyzed using a 600 l/mm grating coupled to a cooled CCD detector (Andor DU420A-BU). In this configuration, only the sample moves during the measurements in order to assure the same illumination condition at each points and the same alignment of the tip apex with the focal spot of the objective. The main advantage of this experimental device is that it allows measurements with any kind of samples (transparent or not) compatible with AFM.

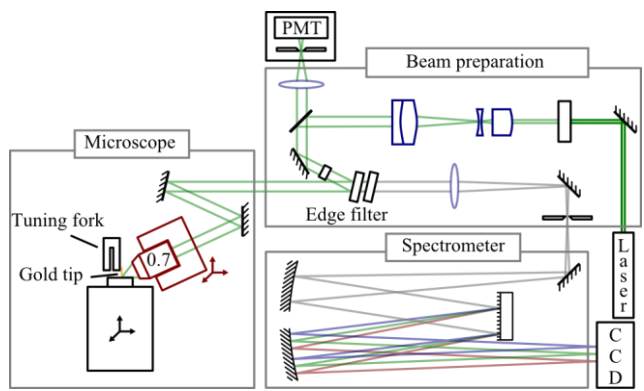

*Figure 1: Experimental setup of the near field scanning optical microscope. It is based on a Tuning Fork AFM optically coupled to a laser excitation and a spectrometer through a side illumination.*

Tips were fabricated by an electrochemical process. A 100 µm gold wire (Goodfellow, Purity 99.99+ Annealed) was immersed by 1 mm into a hydrochloric acid solution (Sigma Aldrich, 37%) and connected as anode to an arbitrary waveform generator (Keithley 3390). The cathode consisted of a gold ring. A pulsed voltage (7.5 V peak to peak, 30 µs pulses, 3 kHz) was applied to the tip with an offset of 0.5 V. Then, the tip was cut to a length of about 300 µm and glued to a tuning fork (Abracon Corporation, AB38T-32.768KHZ) using a UV-curable glue. This process gives smooth tip with a curvature radius down to 10 nm and an aspect ratio estimated to 30:1when the process is highly optimized (see electronic supplementary information for a typical SEM image of a tip).Their final optical properties strongly depend on their exact shape and final size and even a slight variation of these parameters can affect luminescence. Some tips give rise to a very low luminescence whereas some others provide a very intense signal. For our measurement we have privileged the tip with the most intense signal.

Near field measurements were performed on a gold surface (Arrandee™) previously prepared by flame annealing. The optical near field study of such a sample appears as a major challenge: it is not transparent and it has a rather low intrinsic signal compared for instance with plasmonic metallic nanostructures[21,22].

The luminescence peak analysis was performed using homemade software. It allows performing complex mathematical operations such as non-linear regression (using the ALGLIB library) of peaks on complete mapping or algebraic operations on and between maps and spectra.

**Results and Discussion**

In *Figure 2*a we show the luminescence spectra from the apex of the tip when the tip is not in contact with the gold surface and when it is in contact. The intensity of this luminescence is clearly higher at the apex of the tip as can be seen in the mapping shown in *Figure 2*b and c. A strong luminescence signal has already been reported in previous tip enhanced spectroscopy experiments and was attributed to inelastic light scattering from the adsorbate-substrate system[23]. If it was the case here we should have this signal whatever the tip, and this is not what we observed. We therefore attribute this signal to gold interband transitions enhanced by the tip-sample coupling. In aSNOM the tip can be approximated as a nanorod. The luminescence of gold nanorods has already been studied[24,25]: it can be extremely intense compared to a gold surface. The final enhancement and the emission wavelength strongly depend on the shape of the rod; this is surely the reason why luminescence is not always detected even with our tips. In any case, the final intensity of this luminescence is a result of the strong near field enhancement of the incoming and outgoing light via the excitation of the surface plasmon resonance of the tip itself. Several numerical studies have shown that this enhancement can be affected by different parameters, notably the environment of the plasmonic nanostructure. In *Figure 2*a it appears clearly that luminescence of the tip is enhanced when it is in contact with the gold surface. This is induced by the strong tip-surface coupling[26]. Numerical simulations have conclusively shown that when a plasmonic tip is very close to a gold surface a tremendous near field enhancement is observed near the apex[27,28]. This very enhancement is used in tip-enhanced spectroscopy. In our case, this enhancement increases the intensity of the luminescence. This feature clearly confirms that there is a strong optical coupling between the tip and the gold surface.

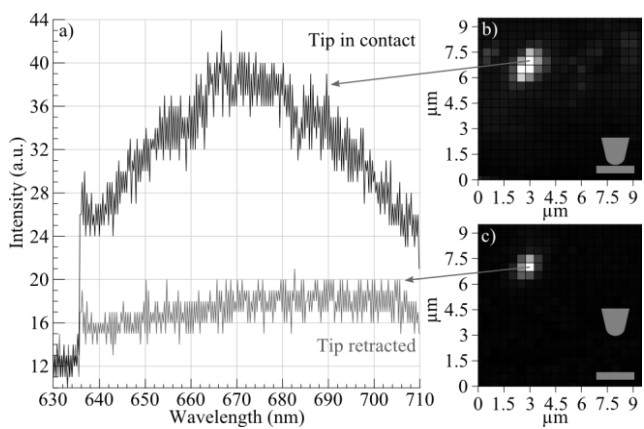

*Figure 2: a) Luminescence from the apex of the gold tip. b) and c) Mapping of the luminescence intensity of the tip when it is respectively in or not in contact with the gold surface. Insets represent the tip (top) sample (bottom) configuration seen by the scanning objective.*

The most commonly used analytical model for the description of the tip sample interaction is the dipole model. The tip is approximated by a sphere (with a radius equal to the curvature radius of the tip). The tip-sample optical coupling is seen as the interaction between this sphere and its image due to the substrate[29]. More sophisticated and realistic models have been developed since [11] and the full interpretation of the tip-surface coupling must take into account several parameters: retardation, multipoles excitation, multiple scattering *etc.*[12] In any case, numerical studies have shown that tip-sample coupling sensitively depends on both the tip and sample materials as well as the ambient medium. For linear light scattering the optical image is based on a dielectric contrast[30]. In *Figure 3*a and b we show the topography and the corresponding elastic light scattered mapping of a gold surface. Even if some similarities are clearly visible, a comparison of both images reveals different information. In the optical image, some details with a typical dimension of a few 10 nm are visible. This confirms the high spatial resolution of optical near field measurements. The final contrast of this image is a complex convolution of topography and optical properties including local and propagating plasmons of the surface. It needs to be said that several studies have shown that the interpretation of a SNOM image is indeed far from obvious and that knowledge of even hardly accessible local materials parameters is required to avoid ambiguities[31,32]. The most common artifact is induced by the strong dependence of the signal on the tip-sample distance. This explains why the final image is a complex convolution of topography and optical properties of the surface. The deconvolution between those two contributions is beyond our present knowledge of local information. In addition, in most of SNOM studies based on the elastic scattered signal the final contrast is very weak and different experimental techniques must be used to improve it (harmonic demodulation and/or interferometric measurements). In our experimental device, none of these methods is required. But it explains why the contrast in *Figure 3*b is relatively weak and that some details are not visible.

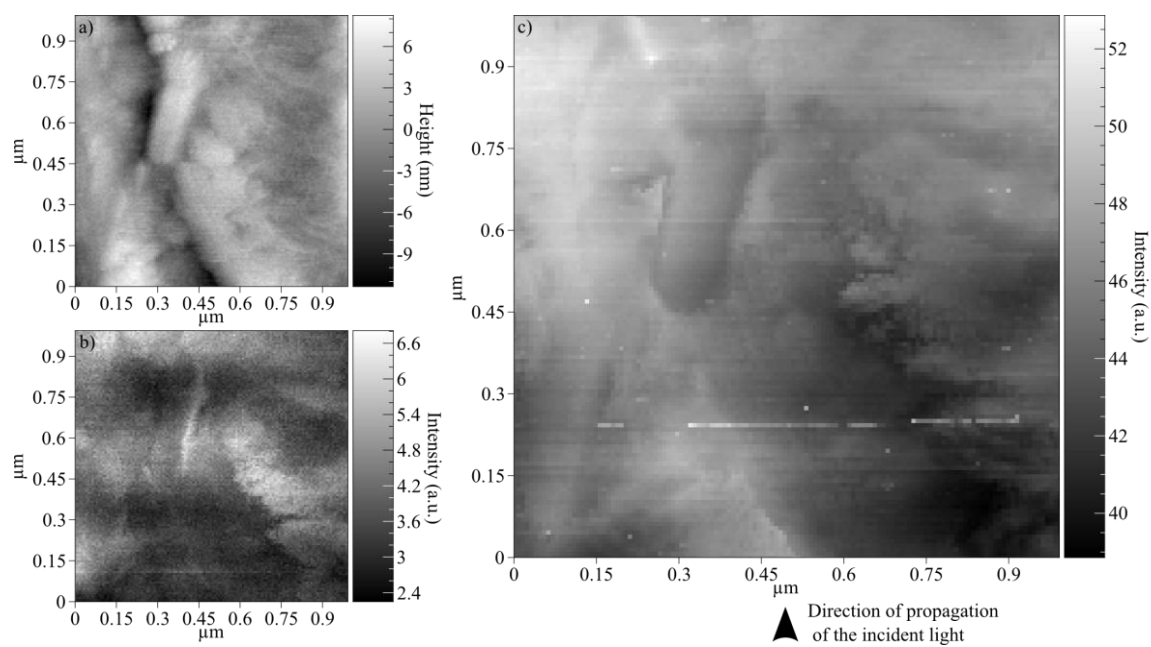

*Figure 3: 1 µm × 1 µm scanning images of the gold surface a) Topography b) Intensity of the elastically scattered light c) Intensity of the luminescence.*

To improve these limitations we have recorded the same optical image using the luminescence signal. The corresponding mapping can be seen in *Figure 3*c. We would like to point out that, as the luminescence signal is very intense the acquisition time for each point has been limited to 100 ms, the time cost for acquiring a 128 × 128 image is thus less than one hour. It offers an excellent compromise between spatial resolution and limited drift of the piezoelectric controller. The obtained image offers strong similarities with both the topography (*Figure 3*a) and optical properties (*Figure 3*b). Some details not visible in *Figure 3*b can now be clearly distinguished. The very high spatial resolution of the optical image is remarkable. Some spatial details that are hardly visible with the AFM topography image appear clearly on the optical image. This result confirms that the final resolution of a SNOM image can be higher than the curvature radius of the probe[33]. Cross sections (See electronic supplementary information) reveal that the spatial resolution of the optical image is typically around 20 nm.

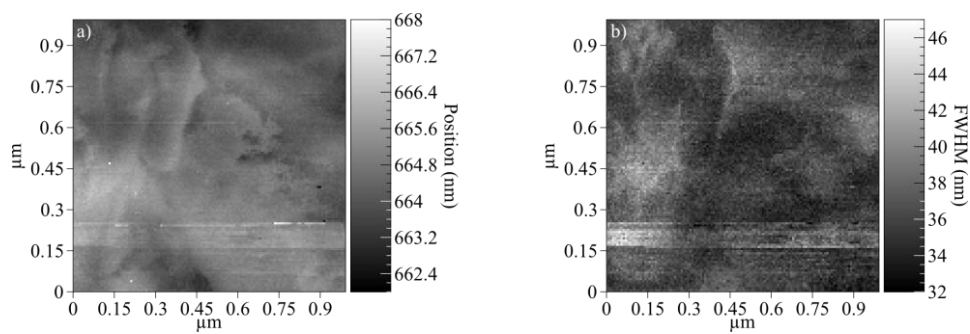

*Figure 4: 1 µm × 1 µm scanning images of the gold surface. The luminescence curve is fitted with a lorentzian and the following parameters are reported a) Position of the maximum b) Full Width Half Maximum map of the fitted curve.*

As evocated above the *Figure 3*b is dominated by topographic artifacts and surely contains little optical information. The fact that *Figure 3*c is different suggests that this image is now dominated by the optical contribution. To confirm this hypothesis we show in *Figure 4*a and b respectively the position of the luminescence maximum and the full width half maximum (FWHM). Both parameters were obtained fitting the curves of each point with a lorentzian with our homemade program. The luminescence of a nanorod, as any plasmonic nanostructure, is strongly influenced by the optical properties of its environment. As a consequence, we derive that a combined evolution of intensity, position and FWHM of luminescence is a sign of an optical interaction between the probe and the optical properties of the sample. This is exactly what we observe here: *Figure 3*c, *Figure 4*a and b clearly show strong similarities. Another experimental result demonstrates that those images are dominated by the optical contribution. In *Figure 4* the region between y = 0.15 µm and 0.25 µm is clearly brighter than the rest of the image. It appears that the luminescent wavelength is slightly higher (the typical shift is a few nanometers) and the FWHM has increased. We have observed that for this region some Raman modes are present (see *Figure 5*).

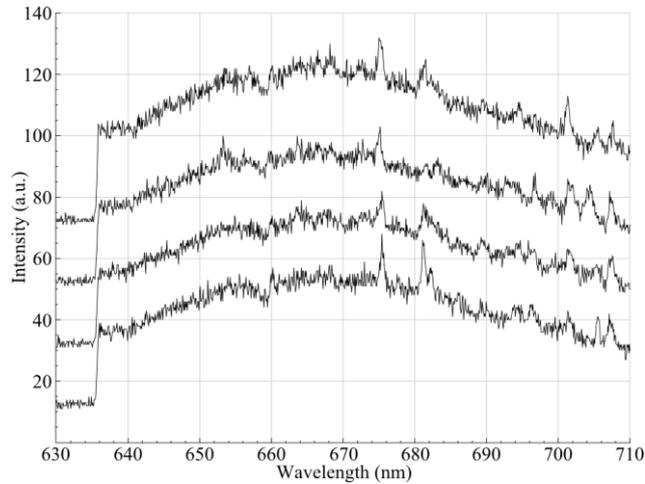

*Figure 5: Luminescence for the region at y = 0.25 µm. Some unidentified Raman modes (sharp peaks) are clearly present. The offset between the four spectra is for clarification only.*

They originate from organic impurities that were temporarily adsorbed at the tip apex and therefore modify the dielectric medium around the tip. No Raman modes are observed outside this region. The Raman modes of these parasitic molecules are also strongly enhanced by the near field antenna as for any TERS measurement. Under ambient conditions and considering that the tip represents an optical tweezer, it is almost impossible to prevent such pollution. Among all the parameters evocated above that can affect the nanorod luminescence only one can explain this result: a change in the optical index in the medium surrounding the nanorod. As the amount of deposited molecules is very low the induced index change is certainly very weak, yet it is clearly visible in *Figure 4*. This result demonstrates the high optical sensitivity of the luminescence and that the corresponding image is dominated by the optical contribution compared to topographic artifacts. As for any SNOM measurements, this optical signal is the result of two different contributions: the optical properties of the sample itself, that is to say a plasmonic surface, and the optical coupling between the tip and the sample. It is extremely difficult to determine the exact contribution between those two features. Pettinger et al.[23] have studied the evolution of luminescence background in TERS measurements. They reported a blue shift and an intensity decrease when the tip-sample distance increases and attributed it to a decreasing electrodynamic coupling between the excited surface plasmons of the tip and the gold surface. Even if we cannot totally exclude this contribution, we do not consider it to be dominant in our measurements for a simple technical reason: each spectrum is acquired during 0.1 s, a very long time compared to what is required for the feedback loop to insure a constant tip-sample distance. As a consequence we assume that the main features observed in both *Figure 3* and *Figure 4* are a sign of the optical properties of the gold surface. Those properties are dominated by the field distribution around plasmonic nanostructures. Such nanostructures exhibit strong near field enhancement at their edge. This is what we observe in *Figure 3*c: the luminescence intensity is higher at the edge of the gold nanostructures. As a consequence, it appears that this image is the distribution of the electromagnetic field on the gold plasmonic surface.

As a conclusion we suggest that, even if the presence of a tip can affect the optical properties of the sample[34], our luminescence-based near field mapping technique is a simple and very efficient tool for the study of the optical properties of surfaces. Our measurements have evidenced several advantages of this method: a very high signal to noise ratio leading to reasonable acquisition time compatible with high resolution scanning, a very high sensitivity of the luminescence parameters with the optical environment of the tip, a simple experimental device without complex post analysis and a technique that can be used with any kind of sample compatible with AFM measurements. Our study has demonstrated that this approach is extremely suitable for the study of plasmonic surfaces. We suggest that it should be applied for the optical study of nanoantennas[35], an extremely promising and growing field in nano-optics and surface enhanced spectroscopy.

**Acknowledgements:** A.M. acknowledges the Agence Nationale de la Recherche Scientifique (ANR) for financial support under the CARIOCA project (2010-JCJC-918-01). A.R. is grateful for a CFI leaders opportunity fund for his infrastructure as well as an NSERC discovery grant and an FRQNT team grant.